\documentclass{article}
\usepackage{amsmath}
\usepackage{amssymb}
\begin{document}
\def\r{{\bf r}}
\def\x{{\bf x}}
\def\y{{\bf y}}
\def\t{{\mathbf\theta}}
\def\o{{\mathbf\omega}}
\def\s{\sigma}
\def\a{{\bf a}}

\def\ket#1{\vert#1\rangle}
\def\bra#1{\langle#1\vert}
\def\ave#1{\langle#1\rangle}
\def\half{{\scriptstyle{\frac{1}{2}}}}
\def\ihalf{{\scriptstyle{\frac{i}{2}}}}

\title{Notes on Certain Newton Gravity Mechanisms of
Wave Function Localisation and Decoherence}
\author{Lajos Di\'osi\\
Research Institute for Particle and Nuclear Physics\\
H-1525 Budapest 114, POB 49, Hungary}
\maketitle

\begin{abstract}
Both the additional non-linear term in the Schr\"odinger equation and the 
additional non-Hamiltonian term in the von Neumann equation, proposed
to ensure localisation and decoherence of macro-objects, resp., contain the same Newtonian
interaction potential formally. We discuss certain aspects that are
common for both equations. In particular, we calculate the enhancement
of the proposed localisation and/or decoherence effects, which would take
place if one could lower the conventional length-cutoff and resolve the 
mass density on the interatomic scale.
\end{abstract}

\section{Introduction}\label{Introduction}
\begin{figure}\label{Fig}
\begin{center}
\setlength{\unitlength}{0.025in}
\begin{picture}(60,60)(0,-10)
\put(0,0){\line(1,0){60}}
\put(0,0){\line(3,4){30}}
\put(60,0){\line(-3,4){30}}
\put(-30,0){$i\hbar\dot\Psi=H\Psi$}
\put(65,0){$\Delta\Phi=4\pi Gf$}
\put(5,42){$c^2t^2-r^2=\mathrm{invariant}$}
\put(5,2){$\hbar$}
\put(50,2){$G$}
\put(29,34){$c$}
\put(12,-5){von Neumann}\put(29,-10){?}
\put(-10,25){Dirac}\put(-10,20){positron}
\put(50,25){Einstein}\put(50,20){black hole}
\put(20,15){$H\Psi=0$}
\end{picture}
\end{center}
\caption{Scheme of Physics' Building. $c=$velocity of light, $G=$Newton's
gravitational constant, $\hbar=$Planck constant. The corners of the triangle
represent the three fundamental theories, the sides correspond to partially
unified theories while the middle symbolises the fully unified theory.}
\end{figure}
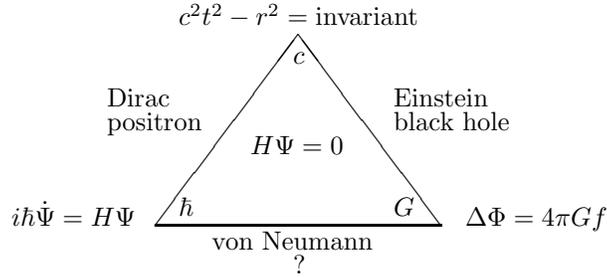

Experts in history of science may perhaps know what von Neumann's approach
would be to the concept of a fully quantised Universe. His measurement
theory yields perfect statistical interpretation of the quantum state
as long as there exists a classical --- non-quantised --- sector of the Universe. 
The challenge of a fully quantised Universe has been attracting many theorists 
even in the lack of pressing experimental evidences. 
Where might those evidences --- or at least indications --- come from? 
That must be the combination of extreme high energies and extreme high 
gravitating mass densities. As a consequence, the mainstream concept of a 
quantised Universe targets a quantised cosmology through the quantisation of the Einstein
theory of space-time. Despite theoretical efforts through the past decades, that big step has 
not been done so far. Experts do not agree what the bottle-neck is. 
It may be our concept of the quantum or our concept of the space-time.
Both, certainly. I used to emphasise one: the bottle-neck is the quantum. 
The von Neumann theory of measurement becomes useless
if the whole Universe is quantised. To make a shortcut to our subject,
we cite a figure from ref.~\cite{Dio92}, with the Schr\"odinger equation
of the Universe written in the middle, see fig.~\ref{Fig}. 
Our failure to interprete the universal wave 
function $\Psi$ may not be related to relativity. The formal argument of the 
figure is almost categoric: one of the three partially unified theories
is missing. Then, why not, the bottle-neck may be the missing unified theory 
for quantum mechanics $(\hbar)$ and Newtonian gravity $(G)$ --- once called 
Newtonian Quantum Gravity. 
We may assume that the path upto a relativistic theory of a quantised 
Universe goes through the non-relativistic theory of Newtonian Quantum 
Gravity explaining the quantised motion of common macroscopic objects.  
In particular, we can make a small step toward the theory of quantised
Universe if we establish a theory of ``spontaneous'' measurement of quantised
non-relativistic macro-objects. 

For the past 20 years, many authors have considered the possible role of 
Newtonian gravity in resolving the apparent controversy between 
the common classical motion of macro-objects and their quantum mechanical
description \cite{Dio92}-\cite{Ges04}. 
I will focus on specific old proposals 
where the standard Schr\"odinger-von-Neumann equations of quantum mechanics 
are modified by
concrete gravitational terms of simple and transparent mathematical structure.

\section{Two mechanisms, two models, one Newtonian structure}
\label{Two}
The studies of our interest concentrated on two inter-related elements of 
classical behaviour of a rigid macro-object: precise center of mass 
{\it localisation} and {\it decoherence} (decay) of superposition 
between separate positions.
To guarantee the first, the attractive Newtonian self-consistent
gravitational field was introduced into the Schr\"odinger equation
\cite{Dio84,MorPenTod98}.
To guarantee the second, a universal decay mechanism was postulated
for superpositions between separate positions, scaled by the difference
between the corresponding Newtonian field strengths 
\cite{Dio87,Dio89,Pen96}. In both {\it localisation}
and {\it decoherence} mechanisms, resp., 
the relevant quantity is the Newtonian interaction
\begin{equation}\label{UXXpr}
U(X,X')=
-G\int\frac{f(\r|X)f(\r'|X')}{\vert\r'-\r\vert}d\r d\r'
\end{equation}
between two mass densities corresponding to two configurations
$X,X'$ of the macro-objects that form our quantum system. 
Typically for rigid objects, position $X$ contains the center of mass 
coordinates $\x_1,\x_2,\dots$ 
and the rotation angles $\t_1,\t_2,\dots$. For simplicity, we shall consider
spherically symmetric or point-like objects, to discuss their translational 
degrees of freedom. Hence $X$ stands for $\x_1,\x_2,\dots$ only.

With the help of the interaction potential (\ref{UXXpr}), we construct 
the Schr\"odinger-Newton equation for the wave function $\psi(X)$ of
the massive objects \cite{Dio84,MorPenTod98}:
\begin{equation}\label{SchN}
i\hbar\frac{d\psi(X)}{dt}=
\hbox{standard q.m. terms}+\int U(X,X')\vert\psi(X')\vert^2dX'~\psi(X)~.
\end{equation}
The second term on the rhs leads to stationary solitary solutions.
The Schr\"o\-din\-ger-Newton eq. ensures the
stationary localisation of the objects. Yet, the equation can not account for
the expected decoherence of macroscopic superpositions like $\ket{X}+\ket{Y}$.

An alternative, irreversible, equation serves this latter purpose. 
We start from the von Neumann equation which is equivalent with the standard
Schr\"odin\-ger equation. It evolves the density matrix $\rho(X,Y)$
rather than the wave function $\psi(X)$.
The construction of the von-Neumann-Newton equation reads \cite{Dio87,Dio89}:
\begin{equation}\label{vNN}
\frac{d\rho(X,Y)}{dt}=\hbox{standard q.m. terms}
+\frac{U(X,X)+U(Y,Y)-2U(X,Y)}{2\hbar}\rho(X,Y)~.
\end{equation}
The second 
term on the rhs contributes to an exponential decay of the superposition 
$\ket{X}+\ket{Y}$, with the following decoherence time \cite{Dio87,Dio89,Pen96}:
\begin{equation}\label{decay}
\frac{2\hbar}{2U(X,Y)-U(X,X)-U(Y,Y)}~.
\end{equation}

To avoid misunderstandings, we emphasise that the 
Schr\"odinger-Newton eq.~(\ref{SchN})
and the von-Neumann-Newton eq.~(\ref{vNN}) are two {\it alternative} equations 
to modify the standard quantum mechanics for macro-objects. 
In our notes, we shall treat these two separate equations parallel to 
each other because the gravitational terms depend on the same 
Newton interaction (\ref{UXXpr}) in both equations. 
[The desired two effects,
localisation plus decoherence, have been realised in ref.~\cite{Dio89} through
a single stochastic Schr\"odinger/von-Neumann-Newton equation based invariably
on the structure $U(X,X')$.]

\section{Case study of a rigid ball}\label{Ball}
Following tradition, restrict ourselves for the study of a single 
rigid ball of mass $M$ and radius $R$. Its mass density depends 
on the distance from the center of mass $\x$:
\begin{equation}\label{rhorx}
f(\r|\x)=f(\r-\x)~,
\end{equation}
where $f$ is spherically invariant function.
The Newtonian interaction (\ref{UXXpr}) depends on the distance $\x'-\x$:  
\begin{equation}\label{Uxxpr}
U(\x'-\x)=
-G\int\int\frac{f(\r-\x)f(\r'-\x')}{\vert\r'-\r\vert}d\r d\r'~.
\end{equation} 
Through this section, we assume that the characteristic distances
$\vert\x'-\x\vert$ are small compared to any other relevant
length scales of the problem. 
Then we expand the interaction potential upto the first nontrivial order
in $\x'-\x$ \cite{Dio84}:
\begin{equation}\label{Uxball}
U(\x'-\x)=U_0 + \frac{1}{2}M\omega_G^2 \vert\x'-\x\vert^2~,
\end{equation}
where $\omega_G$ is a certain gravitational frequency of self-interacting bulk
matter \cite{Ges04}.
We can write it into this simple form:
\begin{equation}\label{omegaGball}
\omega_G^2=\frac{4\pi}{3M} G\int f^2(\r)d\r~.
\end{equation}
At constant mass density $\bar f=3M/4\pi R^3$ we obtain: 
\begin{equation}\label{omegaGball0}
\omega_G^2=\frac{4\pi}{3} G\bar f=\frac{GM}{R^3}~.
\end{equation}

As we mentioned in sect.~\ref{Introduction}, 
the Newtonian interaction potential (\ref{UXXpr}) plays the
key role in the proposed mechanisms of localisation (\ref{SchN})
or decoherence (\ref{vNN}) of macro-objects.  
Using the approximation (\ref{Uxball}) for $U(X,X')$, we obtain
the non-linear Schr\"odinger-Newton eq.~(\ref{SchN}) for the wave 
function of the center-of-mass of our ball:
\begin{equation}\label{SchNball}
i\hbar\frac{d\psi(\x)}{dt}=-\frac{\hbar^2}{2M}\Delta\psi(\x)
+\frac{1}{2}M\omega_G^2\vert\x-\ave{\x}\vert^2\psi(\x)~.
\end{equation}
For simplicity, we assumed the absence of external potentials. This
non-linear equation has exactly calculable solitary solutions. 
In the co-moving system, the quantum mechanical mean value $\ave{\x}$ 
is constant and the system becomes isomorfic with a harmonic 
oscillator of frequency $\omega_G$. The width of its localised
ground state is the following \cite{Dio84}: 
\begin{equation}\label{loc}
\sqrt{\frac{\hbar}{M\omega_G}}
=\left(\frac{\hbar^2}{GM^3}\right)^{1/4}R^{3/4}~.
\end{equation}
This could be the natural quantum mechanical localisation of the 
ball. As it is obvious from the eq.~(\ref{SchNball}), nothing prevents
the ball from getting into and then remaining in the superposition of two 
localised ground states that are far from each other.

These ``cat'' states of macro-objects can be excluded from the theory via 
the decoherence mechanism modelled by the von-Neumann-Newton master eq.~(\ref{vNN}).
Applying again the approximation  (\ref{Uxball}) for $U(X,X')$, the 
master equation reduces to the following form \cite{Dio87,Dio89}:
\begin{equation}\label{vNNx}
\frac{d\rho(\x,\y)}{dt}=
\frac{i\hbar}{2M}(\Delta_x-\Delta_y)\rho(\x,\y)
-\frac{1}{2\hbar}M\omega_G^2\vert\x-\y\vert^2\rho(\x,\y)~,
\end{equation}
where $\rho(\x,\y)$ is the density matrix of the center of mass.
This equation implies the decoherence time [cf. eq.~(\ref{decay})] 
\begin{equation}\label{dec}
\frac{2\hbar}{M\omega_G^2}\frac{1}{\vert\x-\y\vert^2}
=\frac{2\hbar R^3}{GM^2}\frac{1}{\vert\x-\y\vert^2}
\end{equation}
for the decay of the superposition $|\x\rangle+|\y\rangle$
\cite{Dio87,Dio89,Pen96}.

Most studies \cite{Dio92}-\cite{Ges04} agree 
that the heuristic mass density, e.g., postulating a bulk homogeneous 
ball, yields plausible localisation (\ref{loc}) and decoherence (\ref{dec}) scales. 
In general, the Newtonian localisation and decoherence can be ignored for atomic systems 
while the quantum dynamics of massive bodies becomes dominated by them. It turns out,
however, that the predicted scales depend on the precise definition of the mass
density $f(\r|X)$.

\section{Point-like objects --- divergence, early cutoff}\label{Pointlike}
The Newtonian self-energy $U(X,X')$ diverges for point-like particles when, e.g.:
\begin{equation}\label{point}
f(\r|\x)=M\delta(\r-\x)~.
\end{equation}
This divergence could paralyse both our localisation and decoherence models 
above. Interestingly, the Schr\"odinger-Newton eq.~(\ref{SchN}) remains regular
for point-like particles as well. 
But the von-Neumann-Newton eq.~(\ref{vNN}) becomes divergent. 
Let us follow the analysis by 
Gian-Carlo Ghirardi, Renata Grassi and Alberto Rimini
\cite{GhiPeaRim90}. The von-Neumann-Newton eq. does not conserve the energy.
The rate of increase of the translational energy for a rigid ball  
can be exactly calculated:
\begin{equation}\label{Egain}
\frac{dE}{dt}=\frac{G\hbar}{2M}\int f^2(\r)d\r=\frac{3}{8\pi}\hbar\omega_G^2~.
\end{equation}
This rate diverges for a point-like object. Comparing the above ``heating rate''
with certain experimental evidences, Ghirardi et al. come
to the conclusion that the cutoff on spatial mass density resolution
must be as early as $a=10^{-5}$cm. (The present author used $10^{-12}$cm,
ignorantly, cf.~\cite{Dio89} and also \cite{Dio05}.) The cutoff can technically be realized by
the corresponding regularisation of the Newtonian kernel $1/r$ or, alternatively,
of the mass density $f(\r|X)$. 

The Ghirardi et al. choice is the smoothened $f(\r|X)$:
\begin{equation}\label{fs}
f(\r|X)=
\left(2\pi a^2 \right)^{-3/2}\int\exp\left(-\frac{1}{2a^2}\vert\r-\r'\vert^2\right)
f_0(\r'|X)d\r'~,
\end{equation}
where $f_0(\r|X)$ is the microsopic mass distribution of the point-like or
extended constituents. Eventually, Ghirardi and co-workers adapted their continuous
spontaneous localisation (CSL) theory to the smoothened mass-density $f(\r|X)$. 
The ``mass-proportional CSL'', cf. e.g. \cite{BasGhi03}, uses the simple contact potential: 
\begin{equation}\label{UXXprGPR}
U_{CSL}(X,X')=-\gamma\int f(\r|X)f(\r|X')d\r
\end{equation}
rather than the original Newtonian version (\ref{UXXpr}). 
In CSL, the strength-parameter $\gamma$ is no longer related to Newton's $G$, 
although $\gamma$ is considered a universal parameter. 
Its ultimate range is under careful investigation by Adler \cite{Adl06}.

\section{Interatomic resolution}\label{Interatomic}
How does the interaction potential (\ref{UXXpr}) change if, 
not imposing the early cutoff $10^{-5}$cm of sect.~\ref{Pointlike}, 
we increase the resolution of the mass density toward the interatomic scales?
For ball geometry, eq.~(\ref{omegaGball}) shows that the gravitational 
frequency $\omega_G$ grows with 
the spatial fluctuations of the mass density. To model the fine-structure  
beyond the constant average mass density $\bar f$, suppose
the ball consists of identical atoms of mass $m$ each.
Assume, furthermore, that the atomic mass is blurred on a certain distance
$\s$. One could take a spatial Gaussian distribution of linear spread $\s$. 
For our purposes, little homogeneous balls of radius $\s$ will suitably
represent the individual atoms. Suppose the scale $\s$ is much smaller
than the interatomic distance, yet much greater than the scale of the
center of mass displacement $\vert\x'-\x\vert\ll R$. Then the total atomic
contribution to the rhs of (\ref{omegaGball}) yields:
\begin{equation}\label{omegaGball1}
\omega_G^2=\frac{4\pi}{3} G\bar f_\s=\frac{GM}{\s^3}~,
\end{equation}
where $\bar f_\s=3m/4\pi\s^3$ is the average density of the blurred
atoms. Comparing this result to (\ref{omegaGball0}),
we can resume that the microscopic resolution of the mass density
{\it enhances} the proposed Newtonian gravitational mechanisms. 
The enhancement can simply be characterised via replacing the Newton 
constant $G$ by the following effective constant $\widetilde G$:
\begin{equation}\label{tGball}
\widetilde G = \frac{\bar f_\s}{\bar f} G~.
\end{equation}
The higher the atomic density $\bar f_\s$ the stronger will be the 
proposed gravitational localisation (\ref{loc}) and decoherence 
(\ref{dec}) effects.

Such enhancement depends on the geometry of the massive object.
If the object is a rectangular slab rather than a ball, the 
the gravitational frequency $\omega_G$ (\ref{omegaGball0}) 
as well as the effective Newton constant
$\widetilde G$ will be re-calculated easily.

\section{Closing remarks}\label{ClosingI}
One witnesses a growing number and variety of proposals
that point toward possible experiments in the near future that will test 
the predicted decoherences at least (see, e.g., \cite{Exp}). 
Testing the proposed spontaneous mechnanisms of 
macroscopic localization is, however, completely out of question (unless 
the test of decoherence is considered an indirect test of localisation 
as well).  
In some cases, the decoherence effects predicted by the Newtonian mechanism  
as in eqs.(\ref{vNN}) and/or (\ref{decay}), 
would be too week to be observed \cite{BerDioGes06}. This tendency may, however, change if
the models resolve the mass density over interatomic scales, see e.g. 
sect.~\ref{Interatomic}, provided the reasons of the earlier length-cutoff 
are somehow neutralised.
We can thus see the issue of mass density resolution is definitive 
from the experimental viewpoint.  

From the theoretical viewpoint, the divergence of the von-Neumann-Newton eq. 
and the corresponding decoherence time for point-like particles represents
a serious issue. Any cutoff turns the parameter-independent model into a less 
attractive one-parameter model. Yet, we do not know whether the Newtonian mechanism, 
i.e. the structure $U(X,X')$, plays a role in spontaneous decoherence or, alternatively,
the simple contact structure $U_{CSL}(X,X')$ of Ghirardi et al. is the real one, while
CSL has been a two-parameter model from the beginning.

If, in the spirit of fig.~\ref{Fig}, we stress that the Newtonian mechanisms are
responsible for the emergent classical behaviour of the massive non-relativistic 
quantised matter then the intellectual perspective includes not only the modification
of the standard quantum mechanics but the refinement of our concept of gravity and
space-time. The Schr\"odinger-Newton and the von-Neumann-Newton eqs. represent the
modification of quantum mechanics.  So far we have not modified or refined our 
concept of gravity. This perspective may overcome the encountered difficulties of the 
the modified quantum equations. It should lead to an autonomous theory of some,
still unknown, new quality of physical phenomena. I wrote this in 1992 and put a
question mark below the edge connecting $\hbar$ and $G$. It marks the radically new
phenomenon that will follow from the autonomous theory --- provided such theory
exists and we discover it. As it happened already for the Dirac and Einstein theories. 
Yet, it is not clear which scenario wins: shall we verify the new physics
through the noise it generates (cf. decoherence) or through the radically new
phenomena that we become able to predict theoretically.

\ \linebreak

This work was supported by the Hungarian OTKA Grant 49384.

\end{document}